\documentclass[prc,reprint,amssymb,aps]{revtex4-2} 

\newcommand{\mc}[1]{\multicolumn{1}{c}{#1}}
\newcommand{\nuc}[2]{$^{#1}$#2}

\usepackage{amsmath}
\usepackage{booktabs}
\usepackage{dcolumn}
\usepackage{diagbox}
\usepackage[pdftex]{graphics}
\usepackage{hyperref}
\usepackage{url}
\usepackage{microtype}
\usepackage{pgfplots}

\hypersetup{
	colorlinks   = true,
	citecolor    = blue,
	urlcolor     = blue,
	linkcolor   = blue,
}

\begin{document}

\title{Doppler-shift attenuation method (DSAM) lifetimes in $^{54}$Cr - a re-evaluation}

\author{A.~E.~Stuchbery}
\affiliation{Department of Nuclear Physics and Accelerator Applications, Research School of Physics, The Australian National University, Canberra, ACT 2601, Australia}

\date{\today}

\begin{abstract}
\begin{description}
\item[Background]
Contemporary stopping powers for $fp$-shell nuclei slowing in tantalum can differ by a factor of two from the 1963 Lindhard, Scharff and Schi{\o}tt (LSS) theory [\href{https://gymarkiv.sdu.dk/MFM/kdvs/mfm\%2030-39/mfm-33-14.pdf}{Mat. Fys. Medd. Dan. Vid. Selsk. 33 no. 14, (1963)}]  used in Doppler-shift attenuation method (DSAM) lifetime measurements dating back to the 1970s. In recent work by Woodside {\em et al.} [\href{https://doi.org/10.1103/16y4-bggw}{Phys. Rev. C 113, 044306 (2026)}], it was found that anomalously high collectivity in the $4^+_1 \rightarrow 2^+_1$ transition of $^{58}$Fe given in the
Evaluated Nuclear Structure Data File [\href{https://www.nndc.bnl.gov/}{ENSDF},
\href{https://doi.org/10.1016/j.nds.2010.03.003}{Nucl. Data Sheets 111, 897 (2010)}] could be traced back as largely due to the use of these historical stopping powers in the Doppler-shift lifetime measurements. Satisfactory agreement with shell-model calculations was obtained from a re-analysis of the 1978 DSAM measurement of Bolotin {\em et al.} [\href{https://doi.org/10.1016/0375-9474(78)90503-1}{Nucl. Phys. A 311, 75 (1978)}] once the stopping powers were replaced by up-to-date values.
\item[Purpose]
The DSAM measurement on $^{54}$Cr by the same group, Stuchbery {\em et al.} [\href{https://doi.org/https://doi.org/10.1016/0375-9474(80)90076-7}{Nucl. Phys. A 337, 1 (1980)}], which used the same experimental methods and procedures, is re-examined.
\item[Method] 
The computer code used in the original DSAM analysis has been rebuilt and upgraded with the capacity to use contemporary stopping powers. Reduced $E2$ transition strengths derived from the revised lifetimes are compared with shell-model calculations.
\item[Results] 
The impact of revised stopping powers on the excited-state lifetimes depends on the relative contributions of nuclear and electronic stopping powers. It is primarily the electronic stopping powers that differ from LSS values. Hence, the magnitude of the change in lifetime depends on the lifetime itself. In the present case, the lifetimes increase by 16\% to 27\%, with the larger increase generally corresponding to shorter lifetimes where electronic stopping dominates.
\item[Conclusions]
 Revised lifetimes, based on current stopping powers, imply $B(E2)$ transition rates between states up to the 6$^+_1$ state in $^{54}$Cr that compare well with shell-model calculations using the GXFP1A interaction. Better agreement is obtained with the effective charges recently proposed by Ogunbeku {\em et al.} [\href{https://doi.org/10.1103/75ry-71sj}  {Phys. Rev. Lett. 135, 072501 (2025)}], namely $e_\pi = 1.3$ and $e_\nu = 0.45$, than with the standard  $e_\pi = 1.5$ and $e_\nu = 0.5$. The lifetime data together with branching ratios and the shell-model calculations strongly suggest revision of the spins assigned to the levels at 3.786 (currently (4)$^+$) and 4.043 MeV (currently 5$^+$). These levels are suggested to be the 5$^+_1$ and 6$^+_2$ states, respectively.
\end{description}
\end{abstract}

\maketitle

\section{Introduction\label{sect:introduction}}

The structure of the $fp$ shell has been the subject of intensive study for many decades. Large-basis shell-model calculations, wherein both protons and neutrons occupy the $0f_{7/2}$, $1p_{3/2}$, $0f_{5/2}$, and $1p_{1/2}$ orbits, are now routine, and can be tested against a considerable body of data. Electromagnetic transition strengths provide a primary set of observables to test shell-model wavefunctions, which are usually based on effective interactions derived from fits to energy levels.

Through the late 1960s to the early 1980s the lifetimes of many excited states in the $fp$ shell were measured by the Doppler-shift attenuation method (DSAM) with the required stopping powers evaluated using the theory of Lindhard, Scharff and Schi{\o}tt (LSS) \cite{Lindhard1963}. It was recently discovered \cite{Woodside2026} that the anomalously strong literature value \cite{ENSDF,Nesaraja2010} for the experimental strength of the $4^+_1 \rightarrow 2^+_1$ transition in $^{58}$Fe could be attributed to the use of LSS electronic stopping powers, which for $^{58}$Fe stopping in tantalum, differ by a factor of two from more recent values given by SRIM \cite{Ziegler2010}. Woodside {\em et al.} \cite{Woodside2026} performed a re-evaluation of the DSAM lifetime measurement on $^{58}$Fe reported by Bolotin {\em et al.} \cite{Bolotin1978}. The LSS stopping powers were replaced by current values from SRIM, which brought the transition strength of the $4^+_1 \rightarrow 2^+_1$ transition into agreement with shell-model calculations and a new measurement by Coulomb excitation \cite{Woodside2026}. This example is unlikely to be unique: all DSAM measurements that used LSS stopping powers should be checked.

Along with the study of $^{58}$Fe, the Bolotin group also performed DSAM measurements on $^{48}$Ti \cite{Linard1978}, $^{50}$V \cite{Kennedy1977}, $^{54}$Cr \cite{Stuchbery1980}, $^{62}$Ni \cite{Kennedy1978}, and $^{63}$Cu \cite{Ryan1980}. A re-evaluation of these measurements to take account of updated stopping powers is recommended. The present paper considers the case of $^{54}$Cr for which the author has access to the original log book.

After a review of the experimental technique in section~\ref{sect:experimental}, the revised data analysis is presented in section~\ref{sect:analysis}. Section~\ref{sect:discussion} begins with a comparison of the new results with previous data, both the previous analysis of the 1980 DSAM experiment and results obtained by other measurements (section \ref{sect:discuss-prevwork}). The experimental data are then discussed in comparison with shell-model calculations in section~\ref{sect:SMcomp}. This comparison suggests a revision of the spins assigned to the levels at 3.786 MeV and 4.043 MeV. 
A summary and concluding remarks follow.

\section{Experimental procedures\label{sect:experimental}}

Experimental procedures were described in the original publication \cite{Stuchbery1980} and references therein \cite{Linard1978,Kennedy1977,Hershberger1969} . The important features are summarized here, along with some corrections and clarifications.

\begin{figure}[t]
\begin{center}
\includegraphics[scale=1.0,angle=0,width= 8.0 cm]{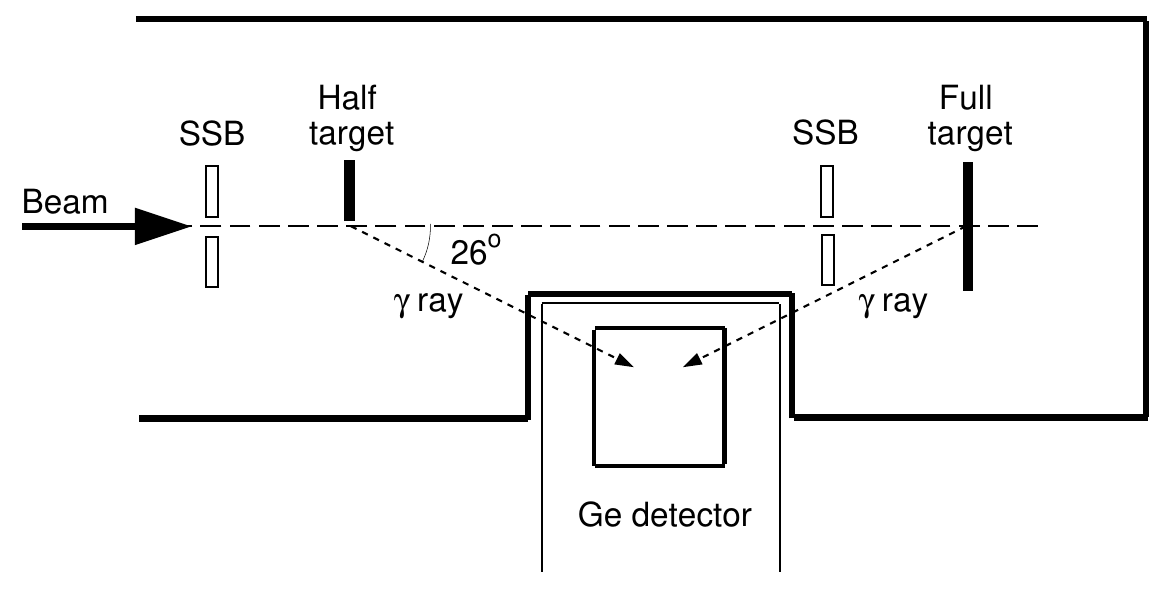}
\caption{Dual target experimental arrangement for the DSAM measurement. There is an annular silicon surface barrier (SSB) detector upstream of each target to detect protons and a $\gamma$-ray detector between the two targets. The outline of the vacuum chamber is indicated. This sketch is based on Fig.~1 of Ref.~\cite{Linard1978} and for clarity is not to scale.}
\label{fig:expt-sketch}
\end{center}
\end{figure}

Figure~\ref{fig:expt-sketch} shows the experimental arrangement.
Beams of 10-MeV $\alpha$ particles were used to populate $^{54}$Cr via the $^{51}$V($\alpha$,p$\gamma$) reaction. Enriched (99.98\%) $^{51}$V targets on both tantalum and molybdenum backings were prepared by vacuum evaporation of V$_2$O$_5$. The thicknesses of $71.3 \pm 3.5$ and $82.7\pm3.3$ $\mu$m/cm$^2$ on tantalum and molybdenum, respectively, were determined by Rutherford backscattering, which also established the ratio of V to O as 1:($1.4\pm 0.1$).

As shown in Fig~\ref{fig:expt-sketch}, the experiment employed two targets, a `half' target upstream and a `full' target downstream. The targets were 15.28 cm apart. A Ge(Li) detector with crystal diameter  4.5 cm and length 4.58 cm was placed equidistant between the two targets facing the beam axis. In other words, the axis of the detector was perpendicular to the beam axis. The face of the crystal was 1.5 cm from the beam axis.

The $\alpha$ beam partially intercepted the upstream half-target while the remainder of the beam proceeded to the full target downstream. Approximately equal beam currents were maintained on the half and full targets. Separate annular silicon surface-barrier (SSB) detectors were placed upstream of each target and recorded protons emitted in the angular range from 153$^{\circ}$ to 174$^{\circ}$. (These angles are correct. Incorrect opening angles for the particle detectors apparently were transcribed from Ref.~\cite{Linard1978} into Refs.~\cite{Bolotin1978,Stuchbery1980,Ryan1980}. However, this transcription error did not affect the data analysis, which used the dimensions of the particle detectors and the distances from their respective targets to determine the acceptance angles.)

The average angles of coincident $\gamma$-ray emission were 26$^{\circ}$ and 154$^{\circ}$ to the beam axis. The initial average $^{54}$Cr recoil velocity
for population of the ground-state was $\beta = v/c =0.0078$, within a cone of half-angle $< 9^{\circ}$.
(The value of $v/c = 0.0068$ given for the initial velocity in Ref.~\cite{Stuchbery1980} corresponds to $ \langle \beta(0) \cos \theta \rangle $, which is defined below.)

Annular tantalum foils 25 $\mu$m thick covered the front faces of the annular SSB detectors to block back-scattered $\alpha$ particles while allowing transmission of protons from the $^{51}$V($\alpha$, p)$^{54}$Cr reaction. Despite the consequent energy loss and straggling, as well as the kinematic broadening due to the finite acceptance angles of the SSB detectors, the proton spectra retained sufficient energy resolution to identify protons associated with the direct population of specific excited states in $^{54}$Cr. See Fig.~1 of Ref.~\cite{Stuchbery1980}.  The possibility of feeding from higher-populated states to the state of interest is eliminated by gating on the appropriate proton group and forming proton-$\gamma$ coincidences.

\begin{figure}[t]
\begin{center}
\includegraphics[scale=1.0,angle=0,width= 7.8 cm]{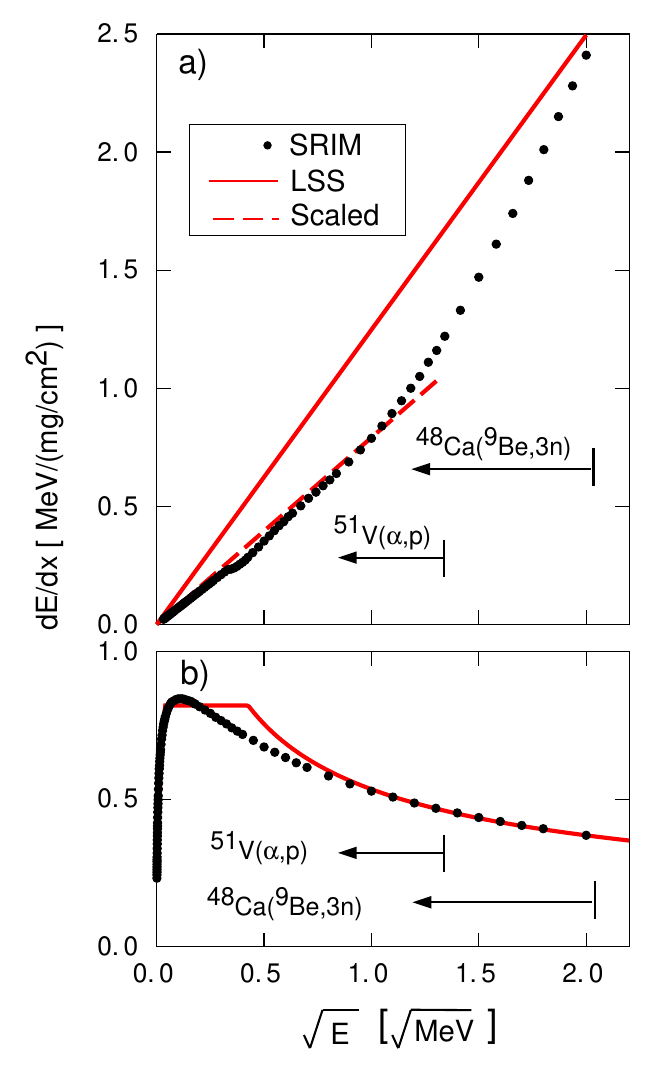}
\caption{
(a) Electronic and (b) nuclear stopping powers for $^{54}$Cr ions stopping in tantalum (100\% $^{181}$Ta) as a function of the square-root of the $^{54}$Cr energy. The LSS stopping powers (solid red lines) are compared with SRIM (black points).
For nuclear stopping the LSS theory is represented by the parametrization introduced by Wozniak, Hershberger and Donahue \cite{Wozniak1969}.
Arrows indicate the energy ranges applicable to the $^{51}$V($\alpha$,p) measurement of Ref.~\cite{Stuchbery1980} and the $^{48}$Ca($^9$Be,3n) reaction of Nathan {\em et al.} \cite{Nathan1978}, which is discussed in sect.~\ref{sect:discuss-prevwork}.
The red dashed line indicates an approximate scaling of the LSS theory to fit the SRIM electronic stopping powers in the regime applicable for the $^{51}$V($\alpha$,p) reaction as used in Ref.~\cite{Stuchbery1980}.
}
\label{fig:dedx-comp}
\end{center}
\end{figure}

\begin{table*}[ht]
	\begin{ruledtabular}
		\caption{
			Lifetimes in $^{54}$Cr from the measurement with the tantalum target backing. Excitation energies $E_x$ and $J^\pi$ assignments are from the current adopted values in Ref.~\cite{ensdf54}, with two exceptions indicated by footnotes. The transition energy $E_{\gamma}$ corresponds to the energy difference between the energies of the initial and final states. Both $E_x$ and $E_{\gamma}$ are rounded to the nearest keV. $\Delta E_{\gamma}$ is the observed energy difference due to the Doppler shift between $\gamma$-rays from the upstream and downstream targets. The experimental attenuation factor is $F(\tau)=\Delta E_{\gamma}/(E_{\gamma} \Delta \langle \beta \cos \theta\rangle$. In the case where two transitions depopulate the state, the average $F(\tau)$ value is denoted in angle brackets. The estimates of the uncertainties in lifetimes due to uncertainty in the electronic and stopping powers are designated $\sigma_{\rm elec}$ and $\sigma_{\rm nuc}$, respectively. $\sigma_{\rm elec}$ corresponds to $(\frac{dE}{dx})_{e} \pm 5\%$; $\sigma_{\rm nuc}$ corresponds to $(\frac{dE}{dx})_{n}\pm 10\%$.
            \label{tab:results-dsamTa}
		}
		\begin{tabular}{dcdddcdcc}
			\mc{$E_x$} & $J^{\pi}$ & \mc{$E_{\gamma}$} & \mc{$\Delta E_{\gamma}$} & \mc{$\Delta \langle \beta \cos \theta  \rangle \times 10^{3}$} & \mc{$F(\tau)$}   & \multicolumn{1}{c}{$\tau$~(ps)}  & \multicolumn{2}{c}{$(\frac{dE}{dx})$ uncertainty~(ps)}  \\ \cline{8-9}
			\mc{(MeV)} &           & \mc{(MeV)}        & \mc{(keV)}   & & & & \mc{$\sigma_{\rm elec}$ } & \mc{$\sigma_{\rm nuc}$} \\ 
			\hline
            0.835   & 2$^+$       &  0.835    &   0.11 \pm 0.17 &   13.53  &    $ 0.009 \pm 0.015 $ & > 4.5 & $\pm 0.34$ & $\pm 1.18$ \\
            \\
            1.824   & 4$^+$       &  0.989    &   0.57 \pm 0.15 &   13.28  &    $ 0.043 \pm 0.011 $ & 4.2^{+1.5}_{-0.9} & $\pm 0.07$ & $\pm 0.24$ \\
            \\
            2.620   & 2$^+$       &  1.785    &  12.57 \pm 0.74 &   13.06  &    $ 0.539 \pm 0.032 $ & 0.20 \pm 0.02 & $\pm 0.00$ & $\pm 0.01$ \\
            \\
            2.830   & 0$^+$       &  1.995    &  11.75 \pm 1.85 &   13.00  &    $ 0.453 \pm 0.071 $ & 0.27^{+0.08}_{-0.06} & $\pm 0.00$ & $\pm 0.01$\\
            \\
            3.074   & 2$^+$       &  2.239    &  28.51 \pm 1.20 &   12.93  &    $ 0.985 \pm 0.041 $ & 0.006^{+0.016}_{-0.006}  \footnotemark[1] \footnotemark[2] & $\pm 0.000$ & $\pm 0.000$ \\
            \\
            3.160   & 4$^+$       &  1.336    &   5.68 \pm 0.49 &   12.91  &    $ 0.329 \pm 0.028 $\\
                    &             &  2.325    &  10.16 \pm 1.24 &          &    $ 0.339 \pm 0.041 $ \\ \cline{6-6}
                    &             &           &              &          &    $\langle 0.332 \pm 0.023 \rangle $ &  0.42 \pm 0.04 & $\pm 0.01$ & $\pm 0.02$ \\
                    \\
            3.222   & 6$^+$       &  1.399    &   4.12 \pm 0.34 &   12.89  &    $ 0.229 \pm 0.019 $ & 0.68^{+0.07}_{-0.06} & $\pm 0.01$ & $\pm 0.04$\\
            \\
            3.393   & $(1^-,2^-)$ &  2.558    &  32.19 \pm 0.91 &   12.84  &    $ 0.980 \pm 0.028 $ & 0.008^{+0.011}_{-0.008}  \footnotemark[1] \footnotemark[3] & $\pm 0.000$ & $\pm 0.000$ \\
            \\
            3.437   & 2$^+$       &  2.602    &  33.45 \pm 0.88 &   12.82  &    $ 1.002 \pm 0.026 $ & < 0.020  & $\pm 0.000 $ & $ \pm 0.000 $ \\
            \\
            3.655   & 4$^+$       &  1.831    &  23.23 \pm 0.31 &   12.76  &    $ 0.994 \pm 0.013 $ &  0.003^{+0.005}_{-0.002} \footnotemark[1] \footnotemark[4] & $\pm  0.000 $ & $\pm 0.000$  \\
            \\
            3.720   & $1^+,2^+$   &  3.720    &  46.00 \pm 2.00 &   12.74  &    $ 0.971 \pm 0.042 $ & 0.012^{+0.016}_{-0.012}  \footnotemark[1] \footnotemark[5] & $\pm 0.000$ & $\pm 0.000$ \\
            \\
            3.786   & $(5)^+$ \footnotemark[6]    &  1.962    &   0.00 \pm 0.60 &   12.72  &    $ 0.000 \pm 0.024 $ &  > 3.7 & $ \pm 0.01 $ & $\pm 0.00 $ \\
            \\
            3.799   & 4$^+$       &  2.964    &  28.10 \pm 0.70 &   12.71  &    $ 0.746 \pm 0.019 $ & 0.092^{+0.008}_{-0.007} & $\pm 0.002$ & $\pm 0.005$ \\
            \\
            4.043   & (6)$^+$ \footnotemark[7] &  0.820    &   8.90 \pm 0.50 &   12.64  &    $ 0.858 \pm 0.048 $ & 0.052^{+0.017}_{-0.016} & $\pm 0.001$ & $\pm 0.002$ \\
            \\

		\end{tabular}
	\end{ruledtabular}
	\footnotetext[1]{Reported previously as a limit.}
    \footnotetext[2]{The 2$\sigma$ limit on $\Delta E_{\gamma}$ gives $\tau <0.037$ ps.}
	\footnotetext[3]{The 2$\sigma$ limit on $\Delta E_{\gamma}$ gives $\tau <0.029$ ps.}
	\footnotetext[4]{The 2$\sigma$ limit on $\Delta E_{\gamma}$ gives $\tau <0.013$ ps.}
	\footnotetext[5]{The 2$\sigma$ limit on $\Delta E_{\gamma}$ gives $\tau <0.042$ ps.}
    \footnotetext[6]{Assigned as (4)$^+$ in the current evaluated data \cite{ensdf54}. The lifetime together with shell-model calculations favors 5$^+$. See sect.~\ref{sect:4043spin}.}
    \footnotetext[7]{Assigned as 5$^+$ in the current evaluated data \cite{ensdf54}. The lifetime together with shell-model calculations favors 6$^+$. See sect.~\ref{sect:4043spin}.}
\end{table*}

The dual-target arrangement effectively doubles the Doppler shift while the average of the up-shifted and down-shifted energies gives the transition energy. Thus, the attenuation factor, which is the ratio of the observed Doppler shift to the maximum Doppler shift, is
\begin{equation}\label{eq:expFtau}
F(\tau) = \frac{\langle E_{\gamma} \rangle_u - \langle E_{\gamma} \rangle_d}{E_{\gamma}^0 ( \langle \beta(0) \cos \theta \rangle_u  - \langle \beta(0) \cos \theta \rangle_d ) },
\end{equation}
where $\tau$ is the meanlife and $\theta$ is the angle between the direction of nuclear recoil and the direction of $\gamma$-ray emission, and the angled brackets represent averages over spatial coordinates and time in the numerator and spatial coordinates only in the denominator. The subscripts $u$ and $d$, respectively, refer to the upstream and downstream targets. The unshifted $\gamma$-ray transition energy is $E_{\gamma}^0 $ and the Doppler-shifted transition energies correspond to $\langle E_{\gamma} \rangle_u$ and $\langle E_{\gamma} \rangle_d$.
Also,
$2 E_{\gamma}^0 = \langle E_{\gamma} \rangle_u +\langle E_{\gamma} \rangle_d$.
The initial velocity of the nucleus of interest relative to the speed of light is $\beta(0)$.

The nuclear lifetime is measured by first determining the experimental value of $F(\tau)$ for the state of interest. The second step is to extract the lifetime by comparing the experimental $F(\tau)$ value with a calculated $F(\tau)$  vs $\tau$ curve. For the evaluation of the experimental $F(\tau)$, $\langle E_{\gamma} \rangle_u$ and $\langle E_{\gamma} \rangle_d$ in  Eq.~(\ref{eq:expFtau}) are simply the Doppler-shifted $\gamma$-ray energies observed in coincidence with the appropriate proton group.

Both the experimental evaluation of $F(\tau)$ and the calculation of the $F(\tau)$ versus $\tau$ curve require the computation of the $ \langle \beta(0) \cos \theta \rangle$  terms in the denominator of Eq.~(\ref{eq:expFtau}). These are obtained by averaging the initial velocity and recoil direction of the $^{54}$Cr nuclei over the acceptance angles of the particle and $\gamma$-ray detectors, taking account of the angle-dependence of the reaction kinematics.

The evaluation of the numerator of Eq.~(\ref{eq:expFtau}) to obtain the $F(\tau)$  vs $\tau$ curve requires the computation of
\begin{equation}\label{eq:Ftau-numerator-av}
\langle E_{\gamma} \rangle = E_0 \langle 1 + \beta(t)\cos \theta(t) \rangle.
\end{equation}
Thus, in addition to the integrals to average over the detection angles and reaction kinematics, there is an averaging over the time-dependence of the $^{54}$Cr velocity and scattering angle as the ions slow to rest \cite{Blaugrund1966}; this is where stopping powers enter into the computation of the $F(\tau)$ curve.

Omitting the spatial averages over the particle and $\gamma$-ray detectors, and changes in the direction of the recoil due to scattering, the theoretical attenuation coefficient is
\begin{equation}
F(\tau) = \frac{\int_0^{\infty} \beta(t)  e^{-t/\tau} dt}{\beta(0) \tau},
\end{equation}
where $t$ is time.
The stopping powers determine $\beta(t)$, the time-dependence of the velocity as the ions of interest slow in the stopping medium. Time, ion velocity, distance travelled, and stopping powers can be related by
\begin{equation}
dt = \frac{dx}{v} = \frac{dE}{v (dE/dx)},
\end{equation}
where $dE/dx$ is the total stopping power, including the electronic and nuclear contributions.
In the nuclear stopping regime, the magnitude of the stopping powers affects the changes in the direction of the recoiling ions due to scattering. Thus, the dependence of the measured lifetime on the stopping powers is both intimate and complex.


For brevity in the tables below, Eq.~(\ref{eq:expFtau}) is rewritten as
\begin{equation}\label{eq:expFtau-short}
F(\tau) = \frac{\Delta E_{\gamma}}{E_{\gamma}^0  \Delta \langle \beta \cos \theta \rangle }.
\end{equation}

Because the $\gamma$-ray emission angles relative to the average recoil direction are near 26$^{\circ}$ and 154$^{\circ}$, $\Delta \langle \beta \cos \theta \rangle \approx 2 \beta(0) \cos26^{\circ} = 0.014$. The actual values are reduced due to the recoil cone of the $^{54}$Cr, and they also decrease with increasing excitation energy.

\begin{table*}[ht]
	\begin{ruledtabular}
		\caption{
			As for Table~\ref{tab:results-dsamTa} but for the measurement with the molybdenum-backed target.
            \label{tab:results-dsamMo}
		}
		\begin{tabular}{dcdddcdcc}
			\mc{$E_x$} & $J^{\pi}$ & \mc{$E_{\gamma}$} & \mc{$\Delta E_{\gamma}$} & \mc{$\Delta \langle \beta \cos \theta  \rangle \times 10^{3}$} & \mc{$F(\tau)$}   & \multicolumn{1}{c}{$\tau$~(ps)}  & \multicolumn{2}{c}{$(\frac{dE}{dx})$ uncertainty~(ps)}  \\ \cline{8-9}
			\mc{(MeV)} &           & \mc{(MeV)}        & \mc{(keV)}   & & & & \mc{$\sigma_{\rm elec}$ } & \mc{$\sigma_{\rm nuc}$} \\

			\hline
            0.835   & 2$^+$       &  0.835    &   0.08 \pm 0.26 &   13.53  &    $ 0.007 \pm 0.023 $ & > 3.8 & $\pm 0.65$ & $\pm 1.49$ \\
            \\
            1.824   & 4$^+$       &  0.989    &   0.81 \pm 0.26 &   13.28  &    $ 0.062 \pm 0.020 $ & 3.3^{+1.6}_{-0.8} & $\pm 0.07$ & $\pm 0.24$ \\
            \\
            2.620   & 2$^+$       &  1.785    &  15.32 \pm 1.14 &   13.06  &    $ 0.657 \pm 0.049 $ & 0.15 \pm 0.03 & $\pm 0.00$ & $\pm 0.01$ \\
            \\
            2.830   & 0$^+$       &  1.995    &  10.28 \pm 3.04 &   13.00  &    $ 0.396 \pm 0.117 $ & 0.38^{+0.23}_{-0.13} & $\pm 0.01$ & $\pm 0.02$\\
            \\
            3.074   & 2$^+$       &  2.239    &  27.40 \pm 1.40 &   12.93  &    $ 0.946 \pm 0.048 $ & 0.024^{+0.020}_{-0.022}  \footnotemark[1] & $\pm 0.001$ & $\pm 0.001$ \\
            \\
            3.160   & 4$^+$       &  1.336    &   6.39 \pm 1.23 &   12.91  &    $ 0.371 \pm 0.071 $\\
                    &             &  2.325    &  11.73 \pm 2.19 &          &    $ 0.391 \pm 0.073 $ \\ \cline{6-6}
                    &             &           &              &          &    $\langle 0.381 \pm 0.051 \rangle $ &  0.40^{+0.09}_{-0.07} & $\pm 0.01$ & $\pm 0.02$ \\
                    \\
            3.222   & 6$^+$       &  1.399    &   5.69 \pm 1.24 &   12.89  &    $ 0.316 \pm 0.069 $ & 0.52^{+0.19}_{-0.12} & $\pm 0.01$ & $\pm 0.03$\\
            \\

		\end{tabular}
	\end{ruledtabular}
	\footnotetext[1]{Reported previously as a limit. The 2$\sigma$ limit on  $\Delta E_{\gamma}$ gives $\tau <0.063$ ps.}
\end{table*}

The following analysis makes use of results reported in Tables 1 and 2 of Ref.~\cite{Stuchbery1980}, supplemented by information from the original analysis logbook. The computer program used in Ref.~\cite{Stuchbery1980} was reconstructed based on optical character recognition of a printed listing.
This code computes the $ \langle \beta(0) \cos \theta \rangle$  and $\langle E_{\gamma} \rangle$ terms in Eq.~(\ref{eq:expFtau}).
After checking that the reconstructed code reproduced the original analysis (both $^{58}$Fe \cite{Bolotin1978}and $^{54}$Cr \cite{Stuchbery1980} measurements), changes were made to examine the effect of modifying the stopping powers, and to automate the procedure for evaluating the level lifetimes from the observed Doppler shifts. Some approximations employed in the original analysis were avoided. Two errors, one in the calculation of energy loss, and one in the evaluation of the average scattering angle, were discovered and corrected; these corrections had negligible effect on the extracted lifetimes. As an improvement, the $F(\tau)$ function was calculated for each excited state, thus taking into account small changes in the \nuc{54}{Cr} recoil velocity.
The original analysis used a single $F(\tau)$ curve for all states due to limited computational resources.
These changes had a small effect on the extracted lifetimes. With the stopping powers as used in Ref.~\cite{Stuchbery1980}, all lifetimes were found to differ very little from those reported therein, and certainly well within the experimental uncertainties.

After checking the veracity of the reconstructed code, it was then modified more extensively to use stopping powers from SRIM \cite{Ziegler2010}. To evaluate the angle changes due to scattering in the nuclear-stopping regime, the code uses the Blaugrund formulation \cite{Blaugrund1966}. This formalism is based on the LSS stopping units, which depend in a complex way on the masses and atomic numbers of both the moving ion and the stopping medium. This dependence complicates the application of the Bragg formulation \cite{Bragg1905,Northcliffe1970} to determine the ``compound" stopping powers (here V$_1$O$_{1.4}$) from those of its elemental components (V and O).
Thus care was taken to ensure that the compound stopping power for the vanadium-oxygen target was treated correctly.

While checking the computer code, it was noted that the integrals over the acceptance angles of the detectors required to evaluate $F(\tau)$ can be replaced by an evaluation at the average detector angles with negligible impact. The kinematics averaging procedures therefore do not introduce any significant uncertainty into the lifetime measurement; the primary sources of uncertainty are the statistical uncertainty in the measured Doppler shift and uncertainties in the stopping powers.

\section{Experimental analysis and results \label{sect:analysis}}

A first assessment of the impact of changes in the stopping power was made by simply scaling the LSS stopping powers to better reflect those given by SRIM.
Figure~\ref{fig:dedx-comp}  compares the LSS stopping powers with those of SRIM for $^{54}$Cr in tantalum. The SRIM electronic stopping powers in the energy range relevant to the measurement are about 60$\%$ of the LSS values.
In contrast, there is better agreement between the SRIM nuclear stopping and LSS the nuclear stopping parametrization of Ref.~\cite{Stuchbery1980}. Overall, the LSS parametrization exceeds SRIM nuclear stopping by a few percent over much of the range of interest.
Under the conditions of the experiment, the slowing of $^{54}$Cr is initially dominated by electronic stopping, but after a time of the order of 10 fs nuclear stopping dominates.

This exercise demonstrated that the mean lifetimes may increase by about 25$\%$ due to the reduced electronic stopping power. The data were then analyzed using the stopping powers from SRIM. Results for the tantalum backing are presented in Table~\ref{tab:results-dsamTa}; those for the molybdenum backing are in Table~\ref{tab:results-dsamMo}.

In these tables the uncertainties given for the lifetimes in the 7th column derive from the energy shift $\Delta E_{\gamma}$ in the 4th column. The effect of a $\pm 5\%$ ($\pm 10\%$) change in the electronic (nuclear) stopping power is indicated in the penultimate (last) column. It is apparent that these uncertainties are small compared with the statistical uncertainty, and become negligible for the shortest-lived states.

It will be noted in Tables~\ref{tab:results-dsamTa} and \ref{tab:results-dsamMo} that most of the short-lived states, which were previously given only as a limit on the mean lifetime, are now given a value, albeit with large uncertainty. The $2\sigma$ limit is also given in a footnote. It was considered reasonable to include lifetime values here rather than limits alone because the reduced electronic stopping powers moved the experimental $F(\tau)$ value to a more sensitive position (longer lifetime) on the $F(\tau)$ versus $\tau$ curve. (Examples of $F(\tau)$ vs $\tau$ curves similar to those that apply here may be found in Fig.~1 of Kennedy {\em et al.} \cite{Kennedy1977}.)

\begin{table}
	\begin{ruledtabular}
		\caption{
			Comparison and combination of lifetimes in $^{54}$Cr determined in tantalum and molybdenum stopping media.
            \label{tab:results-dsam}
		}
		\begin{tabular}{ccccc}
			\mc{$E_x$} & $J^{\pi}$ &  \multicolumn{3}{c}{$\tau$~(ps)}   \\ \cline{3-5}
			\mc{(MeV)} &           &    \mc{Ta backing \footnotemark[1]}      &  \mc{Mo backing \footnotemark[1]}             &    \mc{Adopted \footnotemark[2]}   \\

			\hline
            1.824   & 4$^+$       & $4.2^{+1.5}_{-0.9}$        &   $3.3^{+1.6}_{-0.8}     $  & $3.8^{+1.1}_{-0.7} $     \\
            \\
            2.620   & 2$^+$       & $0.20 \pm 0.02     $       &   $0.15 \pm 0.03         $  & $0.185 \pm 0.025   $      \\
            \\
            2.830   & 0$^+$       & $0.27^{+0.08}_{-0.06}$     &   $0.38^{+0.23}_{-0.13}  $  & $0.28^{+0.08}_{-0.06} $   \\
            \\
            3.074   & 2$^+$       & $0.006^{+0.016}_{-0.006}$  &   $0.024^{+0.020}_{-0.022}$ & $0.010^{+0.013}_{-0.006}$ \\
            \\
            3.160   & 4$^+$       & $0.42 \pm 0.04 $           &   $0.40^{+0.09}_{-0.07}  $  & $0.42 \pm 0.04 $   \\
            \\
            3.222   & 6$^+$       & $0.68^{+0.07}_{-0.06} $    &   $0.52^{+0.19}_{-0.12}  $  & $0.66^{+0.08}_{-0.07} $   \\
            \\

		\end{tabular}
	\end{ruledtabular}
	\footnotetext[1]{Statistical uncertainties are shown as in Tables \ref{tab:results-dsamTa} and \ref{tab:results-dsamMo}.}
    \footnotetext[2]{Average values, including those with asymmetric uncertainties, were evaluated with the Weighted Average option of the Visual Averaging Library \cite{VisAvLib}. Uncertainties include an estimate of uncertainties in the stopping powers (electronic $\pm 5\%$ and nuclear $\pm 10\% $).}

\end{table}

\begin{table}[ht]
	\begin{ruledtabular}
		\caption{
			Revised and original DSAM lifetimes in $^{54}$Cr from Ref.~\cite{Stuchbery1980} compared with other measurements. Nathan {\em et al.} \cite{Nathan1978} also used DSAM. The measurements of Lieb {\em et al.} \cite{Lieb1988}, Kovalenko {\em et al.} \cite{Kovalenko1991}, and Kuronen {\em et al.} \cite{Kuronen1992} were all based on the GRID method. The values given by  Kuronen {\em et al.} \cite{Kuronen1992} are a re-analysis of the measurement by Lieb {\em et al.} \cite{Lieb1988}.
Uncertainties are shown in the abbreviated format as used in ENSDF and Nuclear Data Sheets \cite{ENSDF,ensdf54}. Unless otherwise indicated $J^{\pi}$ is from the evaluated data  \cite{ensdf54}.
            \label{tab:results-omparison}
		}
		\begin{tabular}{lclll}
			\mc{$E_x$} & $J^{\pi}$ &  \multicolumn{3}{c}{$\tau$~(ps)}  \\ \cline{3-5}
			\mc{(MeV)} &           & \mc{Revised} & \mc{Ref.~\cite{Stuchbery1980}}  & \mc{Others} \\
			\hline

            \\
            1.824      & 4$^+$     &  $3.8^{+11}_{-7}     $ & $ 3.5^{+17}_{-11}$    & $2.8(8) $ \cite{Nathan1978} \\
            \\
            2.620      & 2$^+$     & $0.185(25)           $ & $ 0.16^{+4}_{-3}$ & $0.21(6) $ \cite{Lieb1988} \\
                       &           &                           &                         & $0.122(22)$ \cite{Kuronen1992} \\
            \\
            2.830      & 0$^+$     & $0.28^{+8}_{-6}   $ & $0.22^{+9}_{-6} $  \\
            \\
            3.074      & 2$^+$     & $0.010^{+13}_{-6}$  & $< 0.025 $              & $ 0.013(2)$ \cite{Lieb1988} \\
                       &           &                           &                         & $ 0.0031(15)$ \cite{Kovalenko1991} \\
                       &           &                           &                         & $0.0103(5)$ \cite{Kuronen1992}\\
            \\
            3.160      & 4$^+$     &  $0.42(4) $           & $0.35^{+7}_{-6}$   \\
            \\
            3.222      & 6$^+$     & $0.66^{+8}_{-7}   $ & $0.57^{+12}_{-10} $ & $0.7(2)$ \cite{Nathan1978} \\
            \\
            3.393   & $(1^-,2^-)$  & $0.008^{+11}_{-8}$ & $< 0.027 $              & $ 0.021^{+20}_{-10} $ \cite{Kovalenko1991}\\
            \\
            3.437      & 2$^+$     & $< 0.020                $ & $< 0.015$               & $ 0.011(4) $ \cite{Kovalenko1991} \\
            \\
            3.655      & 4$^+$     & $0.003^{+5}_{-2}$ & $< 0.008 $  \\
            \\
            3.720   & $1^+,2^+$    & $0.012^{+16}_{-12}$ & $< 0.043 $              & $ 0.031(5) $ \cite{Lieb1988} \\
                    &              &                           &                   & $ 0.013^{+10}_{-6} $ \cite{Kovalenko1991} \\
                    &              &                           &                   & $ 0.024(2) $ \cite{Kuronen1992} \\
            \\
            3.786      & $(5)^+$ \footnotemark[1]  & $> 3.7                 $  & $> 4 $\\
            \\
            3.799      & 4$^+$     & $0.092^{+8}_{-7}$ & $0.074^{+14}_{-12}$  \\
            \\
            4.043      & (6)$^+$ \footnotemark[1]    & $0.052^{+17}_{-16}$ & $0.041^{+19}_{-15}$ & $ <0.17 $ \cite{Nathan1978} \\
            \\

		\end{tabular}
	\end{ruledtabular}
	\footnotetext[1]{This spin assignment differs from the literature \cite{ensdf54} and is discussed in sect. \ref{sect:4043spin}.}
\end{table}

Table~\ref{tab:results-dsam} compares and combines the results from the tantalum- and molybdenum-backed targets. With the exception of the 4$^+_1$ state, the lifetime data derived from the tantalum-backed target have higher precision. The lifetimes agree within the uncertainties. The results of the two measurements were combined, taking into account the asymmetric errors, using the Weighted Average option of the Visual Averaging Library (V.AveLib) distributed by the International Atomic Energy Agency Nuclear Structure and Decay Data Network \cite{VisAvLib}. As expected, the resultant lifetimes are generally close to those from the higher-statistics measurement with the tantalum-backed target. In the case of the 4$^+_1$ state, however, the lower statistics Doppler-shift measurement from the molybdenum-backed target translates to a lifetime with similar uncertainty to that from the tantalum-backed target because it has a somewhat higher $F(\tau)$ value. The adopted average 4$^+_1$-state lifetime is intermediate between the two values.

The motivation in Ref.~\cite{Stuchbery1980} for taking data on tantalum and molybdenum backings was to test the consistency of the adopted LSS stopping powers for different stopping media. However, the run on the molybdenum-backed target was cut short by an accelerator failure, so no firm conclusions could be drawn. Now, some decades later, with stopping powers based on extensive data \cite{Ziegler2010}, it is considered appropriate to combine the lifetime results from the two data sets.


\section{Discussion} \label{sect:discussion}

\subsection{Comparison with previous work}
\label{sect:discuss-prevwork}

The present reanalyzed DSAM results from Ref.~\cite{Stuchbery1980} are compared with other measurements in Table~\ref{tab:results-omparison}. The previous analysis of the data for the tantalum backing is shown for reference along with measurements by others \cite{Nathan1978,Lieb1988,Kovalenko1991,Kuronen1992}.

Both the present and previous DSAM analysis of Ref.~\cite{Stuchbery1980} data agree with the DSAM lifetimes following the $^{48}$Ca($^{9}$Be,3n$\gamma$) reaction reported by Nathan {\em et al.} \cite{Nathan1978}.  The uncertainties are significant in both measurements. However, because the lifetime of the 4$^+_1$ state currently adopted in the Evaluated Nuclear Structure Data File (ENSDF) \cite{ENSDF} and Nuclear Data Sheets \cite{ensdf54} is from Nathan {\em et al.}, it needs to be recognized that this measurement, like Ref.~\cite{Stuchbery1980}, employed LSS stopping powers and tantalum as the stopping medium. It therefore suffers from the same type of systematic stopping power problem as does Ref.~\cite{Stuchbery1980}. The primary difference between these two measurements is that the heavy ion reaction used by Nathan {\em et al.} resulted in a higher initial recoil velocity for the $^{54}$Cr ions entering tantalum.
This velocity can be estimated to be $\beta \simeq 0.0126$ from the compound nucleus recoil energy, which is $E_i \simeq 4.2$ MeV, or $\sqrt{(E_i)} \simeq 2.05$~MeV$^{1/2}$. In Fig.~\ref{fig:dedx-comp}, it is shown that the SRIM stopping powers pertaining to the measurement of Nathan {\em et al.} are smaller than the LSS values through the energy range appropriate to their measurement.


An estimate of the effect of using SRIM stopping powers in place of LSS was made for the measurement of Nathan {\em et al.} The result is a suggestion that the 4$^+_1$-state lifetime would rise from 2.8 ps to about 3.4 ps, in excellent agreement with the present DSAM value. However, one should be cautious because the uncertainties remain large for both measurements. A new measurement of the 4$^+_1$-state lifetime by the recoil distance method, which is better suited to lifetimes of the order of 3 or 4 ps, should be pursued.

The measurements of Lieb {\em et al.}~\cite{Lieb1988},  Kovalenko {\em et al.}~\cite{Kovalenko1991}, and
Kuronen {\em et al.}~\cite{Kuronen1992} all followed the $^{53}$Cr(n, $\gamma$) reaction. Lieb {\em et al.} used the $\gamma$-ray induced Doppler broadening (GRID) method. Kovalenko {\em et al.} used a variation of this method, but detected the primary $\gamma$~rays in a particular direction such that a Doppler {\em shift} could be observed and the Doppler-shift attenuation method applied.
The lifetimes reported by Kuhonen {\em et al.} constitute a re-evaluation of the GRID results of Lieb {\em et al.}, based on a more sophisticated molecular dynamics simulation of the nuclear recoil.

\begin{figure}[t]
\begin{center}
\includegraphics[scale=1.0,angle=0,width= 7.0 cm]{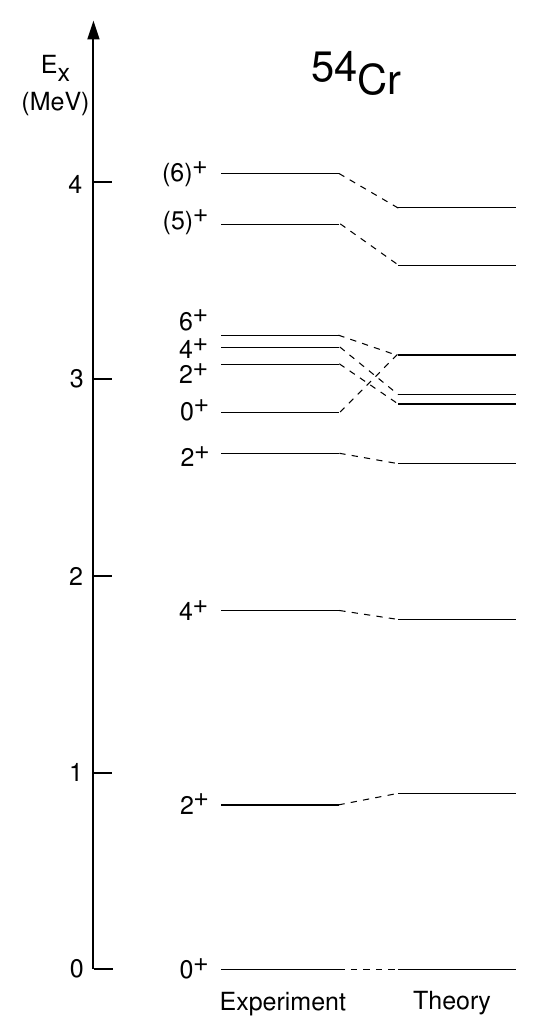}
\caption{
Comparison of experimental and shell-model excitation energies in $^{54}$Cr. The theory uses the complete $fp$ shell and the GXPF1A interactions. Experimental data are from \cite{ensdf54}, apart from the spins of the states shown as  (5)$^+$ and (6)$^+$, which are tentatively assigned in the present work (see  sect.~\ref{sect:4043spin}). All of the experimental and theoretical levels are shown up to and including the 6$^+_1$ state. Above that state only the two higher-spin levels that are suggested to be the 5$^+_1$ and 6$^+_2$ states are shown.
}
\label{fig:levelscomp}
\end{center}
\end{figure}

The measurements following the $^{53}$Cr(n, $\gamma$) reaction generally apply to the shorter lifetimes, near the limit of the DSAM measurements.
Thus, the large uncertainties in the present DSAM analysis generally encompass the GRID-type measurements, which are given with smaller uncertainties. An exception is the comparatively long-lived 2$^+$ state at 2.620 MeV, for which the revised DSAM lifetime has a comparable uncertainty to the GRID measurement and disagrees with it; however, there is agreement with the original value of Lieb {\em et al.} \cite{Lieb1988}. With $F(\tau) \approx 0.54$, the DSAM measurement is near the optimum sensitivity of the $F(\tau)$ curve, whereas the GRID measurement is near its long-lifetime limit. Consequently, the DSAM result should probably be favored in this case.

\begin{table}[t]
	\begin{ruledtabular}
		\caption{Shell-model calculations of $E2$ transition strengths and $Q(2^+_1)$ compared with experiment. Unless otherwise indicated, the $B(E2)$ values are derived from the present lifetime analysis, and where required, branching ratios and mixing ratios are from \cite{ensdf54}. The shell-model calculations designated SM1 and SM2 use the GXFP1A interactions in the full $fp$ shell space. The difference is the use of standard effective charges ($e_{\pi}=1.5$, $e_{\nu}=0.5$) for SM1 and the recently proposed ``universal" effective charges ($e_{\pi}=1.3$, $e_{\nu}=0.45$) for SM2.
            \label{tab:SMcompare}
		}
		\begin{tabular}{ccccccc}
			\mc{$E_i$} & $J_i^{\pi}$ & \mc{$E_f$} & $J_f^{\pi}$ & \multicolumn{3}{c}{$B(E2)$ e$^2$fm$^4$} \\    \cline{5-7}
                       &             &            &             & \mc{SM1}  & \mc{SM2} & Exp. \\

			\hline
            835    & 2$^+$       &  0     &  0$^+$ &   223.5 & 171.1 &  175(7) \footnotemark[1] \\
            \\
            1824   & 4$^+$       &  835   &  2$^+$ &   292.6 & 223.8 &  228$^{+47}_{-51}$      \\
            \\
            2620   & 2$^+$       &  835   &  2$^+$ &    83.6 & 64.1  &  51$^{+29}_{-25}$       \\
                   &             &  0     &  0$^+$ &   0.126 & 0.075 &  1.48$^{+25}_{-21}$ \\
            \\
            2830   & 0$^+$       &  835   &  2$^+$ &    86.8 & 66.6  &  91$^{+24}_{-21}$   \\
            \\                                                          		
            3074   & 2$^+$       &  835   &  2$^+$ &    33.8 & 26.1  &  $<6$ \footnotemark[2] \\
                   &             &  0     &  0$^+$ &    6.38 & 4.74  &  3.3$^{+47}_{-18}$   \\
            \\
            3160   & 4$^+$       &  835   & 2$^+$  &    7.55 & 5.78  &  11.8$^{+17}_{-15}$   \\
                                       \\
            3222   & 6$^+$       &  1824  & 4$^+$  &   258.3 & 197.6 &  233(27)    \\
            \\

            4043   & (6)$^+$ \footnotemark[3]    &  3222  & 6$^+$  &   73.0 & 55.8 &  $88^{+61}_{-36}$ \footnotemark[4]   \\

                   &             &  1824  & 4$^+$  &   44.8 & 34.5 &  $46^{+21}_{-11}$    \\
            \\

\multicolumn{4}{r}{$Q(2^+_1)$ (e fm$^2$):}         & $-26.9$ & $-23.5$ & $ -21(8)$ \footnotemark[1] \\

		\end{tabular}
	\end{ruledtabular}
    \footnotetext[1]{From \cite{ensdf54}.}
    \footnotetext[2]{This is a $2 \sigma$ limit.}
    \footnotetext[3]{ Spin assignment discussed in sect. \ref{sect:4043spin}.}
    \footnotetext[4]{Evaluated with the shell-model value of the mixing ratio, which is $\delta=+0.05$ for both SM1 and SM2.}

\end{table}

\subsection{Comparison with shell model} \label{sect:SMcomp}

In this section, the $E2$-transition strengths that result from the revised lifetimes are compared with shell-model calculations.
Experimental $B(E2)$ values can be evaluated for the decays of all of the excited states up to and including the 6$^+_1$ state.

Shell-model calculations were performed with \textsc{NuShellX}~\cite{Brown2014} in the full $fp$-shell basis space of the $0f_{7/2}$, $1p_{3/2}$, $0f_{5/2}$, and $1p_{1/2}$, orbits for both protons and neutrons, employing the GXPF1A interactions \cite{Honma2004,Honma2005}. The experimental and theoretical level schemes are compared in Fig.~\ref{fig:levelscomp}. These level schemes are complete up to the 6$^+_1$ state. Above this state, however, only the $(5)^+$ and $(6)^+$ states, with proposed revised spin assignments to be discussed in sect.~\ref{sect:4043spin}, are shown.

A number of newer interactions, such as UFPCA \cite{Magilligan2021}, are being developed for neutron-rich nuclei at or near $Z=20$; however,
as a global interaction for the $fp$ shell, the GXPF1A interaction of Honma {\em et al.} \cite{Honma2004,Honma2005} remains appropriate across much of the $fp$ shell. Of particular relevance here, the Cr isotopes $^{48-56}$Cr were included in the fit to binding and excitation energies.

The $B(E2)$ data are presented in Table~\ref{tab:SMcompare}, where they are compared with shell-model calculations performed with two sets of effective charges.

Honma {\em et al.} \cite{Honma2004} used the standard effective charges $\delta e_{\pi} = \delta e_{\nu} =0.5$, or $e_{\pi}=1.5$ and $e_{\nu}=0.5$,  in their survey of electromagnetic properties (i.e. $B(E2)$ values and quadrupole moments) across the $fp$ shell. Calculations with these effective charges are designated SM1 in Table~\ref{tab:SMcompare}. The $E2$ transition rates have been evaluated using harmonic oscillator radial wavefunctions with the oscillator constant taken as $\hbar \omega = 45/A^{1/3} - 25/A^{2/3}$ MeV.
In calculations using the same basis space and the same interactions as used here, Seidlitz {\em et al.} \cite{Seidlitz2011} report $B(E2)$ values about 11\% smaller for the decays of the 2$^+_1$ and 4$^+_1$ states. The difference can be inferred to stem from the use of  $\hbar \omega = 41/A^{1/3}$ MeV in the evaluation of the radial wavefunctions in that work.

Recently, Ogunbeku {\em et al.}~\cite{Ogunbeku2025} proposed the effective charges $e_\pi = 1.3$ and $e_\nu = 0.45$, which they suggested are applicable across both the $sd$ and $fp$ shells. Calculations with these effective charges are designated SM2 in Table~\ref{tab:SMcompare}.

\begin{figure*}[ht]
\begin{center}
\includegraphics[scale=0.95,angle=0,width= 14.0 cm]{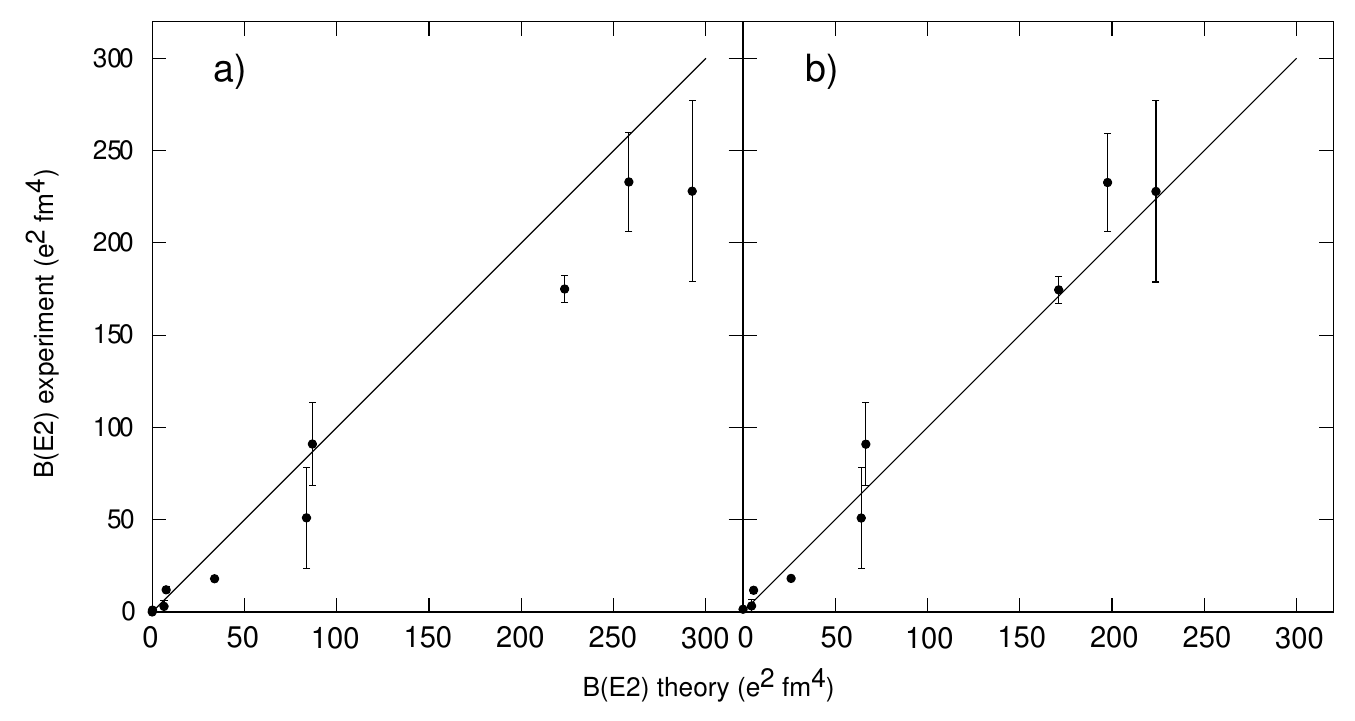}
\caption{Theory versus experiment for $B(E2)$ values between states in $^{54}$Cr up to the $6^+_1$ state.  Standard effective charges, $e_{\pi} = 1.5$ and $e_{\nu} = 0.5$, are employed in panel (a). The recently proposed effective charges of Ogunbeku {\em et al.}~\cite{Ogunbeku2025},  $e_{\pi} = 1.3$ and $e_{\nu} = 0.45$, are employed in panel (b), giving an improved description of the data. See also Table~\ref{tab:SMcompare} wherein SM1 corresponds to panel (a) and SM2 to panel (b).
}
\label{fig:BE2comp}
\end{center}
\end{figure*}

It is evident from Table~\ref{tab:SMcompare}, and the comparisons of theory and experiment visualized in Fig.~\ref{fig:BE2comp}, that the newly proposed ``universal" effective charges better describe the measured reduced transition rates in $^{54}$Cr. This behavior contrasts with that recently found for $^{58}$Fe, where the standard effective charges ($e_{\pi}=1.5$ and $e_{\nu}=0.5$) overall better describe the reduced transition rates of the low-excitation states in that nuclide \cite{Woodside2026}.

The two cases of $^{54}$Cr and $^{58}$Fe are not sufficient to reach firm conclusions, but do suggest that a more comprehensive review of the effective charges applicable for the GXPF1A interaction is called for. It is worth noting here, also, that while it is standard practice to report the effective charges used in shell-model calculations, the value of the oscillator constant used, which also affects the transition strengths, is not always given, but it should be specified.

\subsubsection{Spin-parity assignments for the 3.786-MeV and 4.043-MeV levels}
\label{sect:4043spin}

The comparison of theory and experiment in Table~\ref{tab:SMcompare} and Fig.~\ref{fig:BE2comp} is largely confined to states up to and including the 6$^+_1$ state. Above that state, the density of states increases, experimental spin assignments are missing, and it is harder to make a clear matching between experimental and theoretical levels. There are two experimental levels, however, that can be associated with particular shell-model states, provided that their current spin assignments are revised. These cases are discussed here.

The spin-parity of the 4.043-keV level is assigned as $J^{\pi}=5^+$ in the current nuclear data evaluation \cite{ENSDF,ensdf54}, based primarily on the work of Devlin {\em et al.} \cite{Devlin1999}, which consisted of $\gamma$-ray spectroscopy measurements following the $^{12}$C($^{48}$Ca,$\alpha$n) reaction and shell-model calculations.
The 5$^+$ assignment stems from angular correlation data ((Directional Correlations from Oriented nuclei, i.e. DCO ratios) and comparisons with the shell model using the FDP6 interaction \cite{Richter1991}. The emphasis of the discussion was on revising the previous assignment of $J^{\pi}=7^+$. The present calculations with the GXPF1A interaction agree with the conclusion of Devlin {\em et al.} that no 7$^+$ state is expected near 4-MeV excitation energy. (The yrast 7$^+$ state is predicted at 4.93 MeV.) The possibility that the 4.043-keV state might have $J^{\pi}=6^+$ seems not to have been considered, perhaps because the DCO ratios do not support an $E2$ polarity for the transition to the 4$^+_1$ state. However, it is also noteworthy that Devlin {\em et al.} retain a tentative $J^{\pi}=(5^+)$ in their level scheme.

The calculations with GXPF1A predict the 6$^+_2$ state at $E_x=3.868$ MeV with the 5$^+_1$ state 290 keV lower at $E_x=3.578$ MeV.
If the 4.043-MeV level is in fact the 6$^+_2$ state, then its experimental branching ratios to the 6$^+_1$ and 4$^+_1$ states are well described by the shell model, as is the lifetime of the state.
In contrast, the predicted 5$^+_1$ state has stronger branches to the 4$^+_1$ and 4$^+_2$ states than to the 6$^+_1$ state, which is contrary to experiment if the 4.043-MeV level is the 5$^+_1$ state. Moreover, the shell-model lifetime (taking into account the experimental transition energies) is two orders of magnitude longer than that observed for the 4.043-MeV level.

In Table \ref{tab:SMcompare} the shell-model predicted mixing ratios are used with the experimental lifetime and branching ratios to determine the resultant ``experimental" $B(E2)$ values for decay of the 4.043-MeV level interpreted as the 6$^+_2$ state. These decay properties are clearly in agreement with experiment for both SM1 and SM2. Using the same procedure (i.e. using the shell-model mixing ratios with experimental lifetime and branching ratios) with the 4.043-MeV state identified as the shell-model 5$^+_1$ state, results in unrealistic $B(E2)$ values. For example, the implied $5^+_1 \rightarrow 4^+_1$ transition strength is about 730 e$^2$fm$^4$ or 60 W.u.

Thus, the agreement between theory and experiment is striking if the 4.043-MeV level is the 6$^+_2$ state. This conclusion is contrary to the DCO ratio data \cite{Devlin1999}, however, so further experimentation is required. For now, a tentative assignment of $J^{\pi}=(6)^+$ is proposed.

Having proposed that the 4.043-MeV level is in fact the 6$^+_2$ state, it is natural to ask if one of the observed states can be identified with the expected 5$^+_1$ level. The obvious candidate is the 3.786-MeV level, which currently has $J^{\pi} = (4)^+$ in the evaluated data \cite{ENSDF,ensdf54}, although it is shown as $(4^+,5^+)$ in the decay scheme following $^{54}$V beta decay. The most compelling reason that the 3.786-MeV level may be the 5$^+_1$ state is that experimentally it has a significantly longer lifetime, $\tau > 3.7$~ps, than any of the near-by states, and the only theoretical state in the energy range from 3.0 to 4.5 MeV  with such a long lifetime is the 5$^+_1$ state predicted at 3.578 MeV. Thus, if the 3.786-MeV level is not the 5$^+_1$ state, then its long lifetime is very difficult to explain.

From the shell model perspective, the long lifetime makes the identification of the 3.786-MeV level with the 5$^+_1$ state compelling. Even so, spectroscopic data are required to make a firm spin assignment.

\subsubsection{0$^+$ states} 

Of the calculated levels up to the 6$^+_1$ state, only the 0$^+_2$ state is out of the experimental order, appearing about 300 keV above the experimental level. Nevertheless, the predicted $B(E2; 0^+_2 \rightarrow 2^+_1$) is in good agreement with experiment. Theory predicts the excitation energies for the next two excited 0$^+$ states as $E_x(0^+_3)=3.604$ MeV and $E_x(0^+_4)=4.368$ MeV. The predicted mean lifetimes (taking SM2 effective charges and theoretical energies) are $\tau =79$ fs and $\tau =35$ fs, respectively. There is an experimental 0$^+$ state at $E_x(0^+)=4.013$ MeV, which decays to the 2$^+_1$ state, and matches best with the predicted 0$^+_3$ state. The predicted $(0^+_4)$ state decays primarily to the 1$^+_1$ state. However, the experimental lifetime of the $E_x(0^+)=4.013$-MeV level is of the order of a few femtoseconds \cite{Kovalenko1991}, so if this lifetime measurement is correct, it is difficult to associate the 4.013-MeV level with either of the predicted 0$^+$ states.

A discussion of 0$^+$ states in this region must consider the impact of co-existing deformed states due to excitation of pairs of nucleons across the shell gaps at $Z,N=20$ and $N=28$. For example, $^{40}$Ca has low-excitation deformed and superdeformed 0$^+$ states at $E_x = 2.81$ MeV and $E_x = 6.37$ MeV that are associated with predominantly 4-particle-4 hole and 8-particle-8 hole excitations across $Z=20$ and $N=20$, and which have rotational bands built on them that signal their deformed character. Recent electron-positron pair spectroscopy has shown that these states mix with each other and with the nominally spherical ground state \cite{Ideguchi2022}. This mixing implies that the ground state of $^{40}$Ca is not strictly spherical. Multiparticle-multihole states are found across the Ca isotopes from $^{40}$Ca to $^{48}$Ca. The mixing between the spherical and deformed states in $^{42}$Ca is particularly strong. See Ref.~\cite{Stuchbery2022} section 6, and Figs.~35-36 for a summary of the known 0$^+$ states up to 10 MeV in the even $^{40-48}$Ca isotopes and their associated particle–hole configurations.

\begin{figure}[t]
\begin{center}
\includegraphics[scale=1.0,angle=0,width= 8.5 cm]{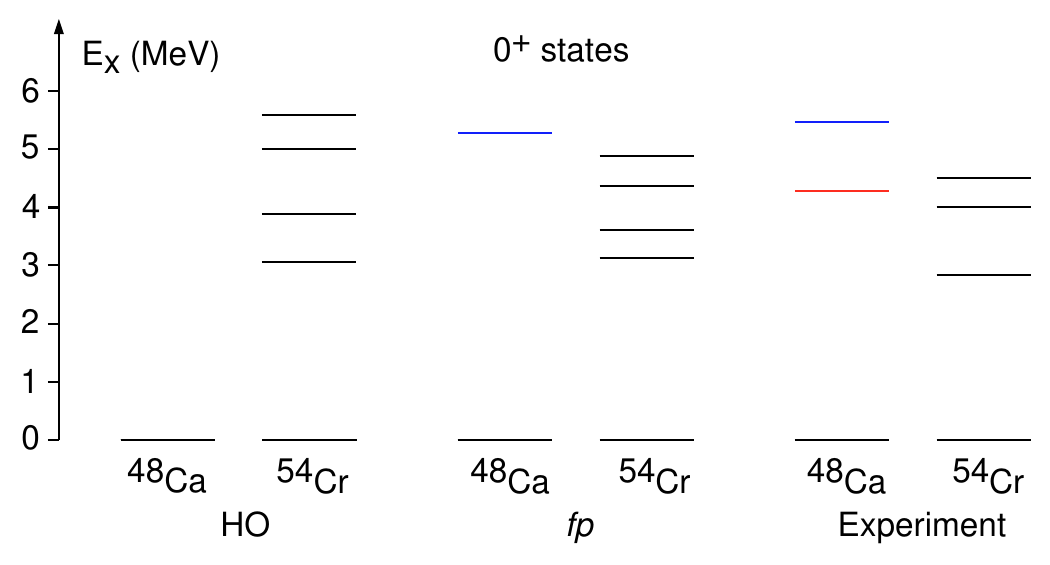}
\caption{Shell-model calculations of 0$^+$ states in $^{48}$Ca and $^{54}$Cr in two basis spaces compared with experiment. The `HO' calculation on the left has a $^{48}$Ca core with protons in  $\pi 0f_{7/2}$ and neutrons in $\nu 1p_{3/2}$, $\nu 0f_{5/2}$, and $\nu 1p_{1/2}$ and the interactions of Horie and Ogawa \cite{Horie1971,Horie1973}. The $fp$-shell calculation in the center uses the full $fp$-shell basis and the GXPF1A interactions as presented in sect.~\ref{sect:SMcomp}. Experimental data are shown on the right. The state near 5 MeV in $^{48}$Ca shown in blue is associated with neutron excitations across $N=28$, which are included in the $fp$ basis space but not in the HO space. The experimental state near 4 MeV in $^{48}$Ca shown in red is associated with proton excitations across $Z=20$, and is outside both basis spaces.
}
\label{fig:0plus-states}
\end{center}
\end{figure}

Of particular relevance here for $^{54}$Cr, which can be considered as the addition of 4 protons and 2 neutrons to $^{48}$Ca, are the excited 0$^+$ states in $^{48}$Ca.
Figure~\ref{fig:0plus-states} compares the excitation energies of 0$^+$ states in $^{48}$Ca and $^{54}$Cr in two shell model calculations and in experiment. To investigate the effect of ``core excitations" in $^{48}$Ca on the 0$^+$ states in $^{54}$Cr, the shell model calculations were performed with different basis spaces. The leftmost calculation in Fig.~\ref{fig:0plus-states}, designated HO, confines protons to $\pi 0f_{7/2}$ and neutrons to $\nu 1p_{3/2}$, $\nu 0f_{5/2}$, and $\nu 1p_{1/2}$. It uses the interactions of Horie and Ogawa \cite{Horie1971,Horie1973}. This basis space has a $^{48}$Ca core, so there are no excited states in $^{48}$Ca. Nevertheless, four excited 0$^+$ states are predicted below 6-MeV excitation energy in $^{54}$Cr.

The second shell-model calculation in Fig.~\ref{fig:0plus-states} uses the $fp$ basis and GXPF1A interaction as discussed in section~\ref{sect:SMcomp} above. There is now a predicted excited state in $^{48}$Ca at 5.275 MeV, which has an experimental counterpart at 5.461 MeV, and is identified with the excitation of a pair of neutrons across $N=28$.
However, there is no theoretical counterpart for the experimental state at 4.284 MeV in $^{48}$Ca, which is identified with the excitation of a pair of protons across $Z=20$, and is thus outside the $fp$ model space. In $^{54}$Cr there are four excited 0$^+$ states predicted below 5 MeV in excitation energy.

Due to blocking effects, excited states associated with both neutron excitation across $N=28$ and proton excitation across $Z=20$ are expected to occur at higher excitation energy in $^{54}$Cr than those  in $^{48}$Ca. The consequence suggested by the trend apparent in Fig.~\ref{fig:0plus-states} is that as the basis space is expanded to include these cross-shell excitations, the excitation energies of the lowest few excited 0$^+$ states in $^{54}$Cr decrease.


Experimentally, there are only three excited 0$^+$ states observed below 5~MeV in $^{54}$Cr. The trends in Fig.~\ref{fig:0plus-states} may suggest that there should be another 0$^+$ state in $^{54}$Cr at an excitation energy of about 3.3 MeV that has not been identified in experiment.
Of the observed states in $^{54}$Cr in this excitation-energy regime, none is a candidate for a 0$^+$ spin-parity assignment.
There may be an as yet unobserved 0$^+$ state, or alternatively, there may be a deficiency in the shell-model calculations. Perhaps, if the missing proton excitation across $Z=20$ were included in the basis space, configuration mixing might sufficiently perturb the 0$^+$ states in $^{54}$Cr to agree with the experimental level scheme, the ``missing" state being pushed to higher excitation energy ($\gtrsim 5$~MeV) where spectroscopic data are scarce.
Further investigation is clearly required.

\section{Summary and Conclusions\label{sect:conclusions}}

Lifetimes in $^{54}$Cr measured by the Doppler-shift attenuation method (DSAM) in Ref.~\cite{Stuchbery1980} have been re-evaluated based on up-dated stopping powers. Whereas the nuclear stopping powers in SRIM \cite{Ziegler2010} differ little from those used in Ref.~\cite{Stuchbery1980}, the updated electronic stopping powers from SRIM are about 60\% of those given by the LSS theory \cite{Lindhard1963} as used in Ref.~\cite{Stuchbery1980}. The consequence is that lifetimes increase by 16\% to 27\%, with the larger increase generally corresponding to shorter lifetimes where electronic stopping dominates. The present lifetimes should replace those from Ref.~\cite{Stuchbery1980} in future evaluations of nuclear data.

The resultant reduced transition rates for states up to the first 6$^+$ state are well described by shell-model calculations in the full $fp$ model space using the GX1PFA interaction \cite{Honma2004,Honma2005} and effective charges of $e_{\pi} = 1.3$ and $e_{\nu}=0.45$, as proposed recently by Ogunbeku {\em et al.} \cite{Ogunbeku2025}. This calculation of the $E2$ transition strengths set the oscillator constant to $\hbar \omega = 45/A^{1/3} - 25/A^{2/3}$ MeV. It is noted that discussions of $E2$ strengths and effective charges should include a specification of the oscillator constant used to evaluate the radial integrals.

Based on comparisons of the shell-model calculations and the decay properties of the excited states in $^{54}$Cr at $E_x = 3.786$ MeV and $E_x = 4.043$ MeV, these states are proposed to be the 5$^+_1$ and 6$^+_2$ states, respectively. Although these assignments are compelling from the shell-model analysis, spectroscopic data are required to determine if these spin assignments are correct.

In older DSAM measurements that used LSS stopping powers the uncertainty in the stopping powers was often assumed to be $\pm 10\%$  \cite{Bolotin1978,Stuchbery1980}. Two cases, $^{58}$Fe \cite{Woodside2026}, and $^{54}$Cr, have now been identified where the electronic stopping powers derived from LSS theory differ by much more than 10\%, in fact up to 50\%, from the more up-to-date values in SRIM. It is clear that such measurements must be evaluated case-by-case, and that repeat measurements, including measurements by alternative methods such as the Recoil Distance Method, may be required in some cases.

\begin{acknowledgments}
This work is supported in part by the Australian Research Council Grants No.\ DP210101201 and DP250100400. I am grateful to Tibor Kib\'edi for discussion and assistance with the software tools distributed by the International Atomic Energy Agency Nuclear Structure and Decay Data Network, and to John Wood for reading and providing useful comments on the manuscript.
\end{acknowledgments}


\bibliographystyle{apsrev4-2} 
\bibliography{cr54-DSAM}

\end{document}